\documentclass[aip,reprint,floatfix]{revtex4-1}
\usepackage{amsfonts,amssymb,amsmath,array}

\usepackage{dcolumn}
\usepackage{bm}

\usepackage[final]{graphicx}
\usepackage{graphicx}
\usepackage{epstopdf}
\graphicspath{{./Figures/}} 

\usepackage[usenames]{color} 

\usepackage{soul} 

\begin{document}


  

\title{Collective effects in confined Active Brownian Particles}

\author{Lorenzo Caprini*}
\affiliation{Scuola di Scienze e Tecnologie, Universit\`a di Camerino, Via Madonna delle Carceri, I-62032, Camerino, Italy}
\author{Claudio Maggi}
\affiliation{NANOTEC-CNR, Institute of Nanotechnology, Soft and Living Matter Laboratory, Roma, Italy}
\author{Umberto Marini Bettolo Marconi}
\affiliation{Scuola di Scienze e Tecnologie, Universit\`a di Camerino, Via Madonna delle Carceri, I-62032, Camerino, Italy}


\begin{abstract}
We investigate a two-dimensional system of active particles confined to a narrow annular domain.
Despite the absence of explicit interactions among the velocities or the active forces of different particles, the system displays a transition from a disordered and stuck state to an ordered state of global collective motion where the particles rotate persistently clockwise or anticlockwise.
We describe this behavior by introducing a suitable order parameter, the velocity polarization, measuring the global alignment of the particles' velocities along the tangential direction of the ring. 
We also measure the spatial velocity correlation function and its correlation length to characterize the two states.
In the rotating phase, the velocity correlation displays an algebraic decay that is analytically predicted together with its correlation length while in the stuck regime the velocity correlation decays exponentially with a correlation length that increases with the persistence time. 
In the first case, the correlation (and, in particular, its correlation length) does not depend on the active force but the system size only. 
The global collective motion, an effect caused by the interplay between finite-size, periodicity, and persistent active forces, disappears as the size of the ring becomes infinite, suggesting that this phenomenon does not correspond to a phase transition in the usual thermodynamic sense.
\end{abstract}

\maketitle


\section{Introduction}

Active matter is comprised of motile active particles that can perform mechanical work at the expense of metabolic or environmental energy 
which drives the system far from equilibrium~\cite{bechinger2016active,marchetti2013hydrodynamics,elgeti2015physics,martin2020statistical}.
Active particles may display fascinating collective phenomena~\cite{gompper20202020}, self-organize in complex spatial patterns and form oriented domains which are responsible for coherent motion~\cite{grossmann2020particle}.
This is the case of flocking birds~\cite{ballerini2008interaction, attanasi2014information} at the macroscopic scale, or the so-called bacterial turbulence, at the microscopic-scale~\cite{wensink2012meso}. 
The latter, characterized by spatial structures in the velocity field~\cite{wioland2016ferromagnetic}, has been originally observed in dense suspensions of \textit{E.~Coli}~\cite{dombrowski2004self} and successively in other species of bacteria~\cite{peruani2012collective, sokolov2009enhanced}.
Large spatial correlations of the velocity field have been also observed in cell-monolayers~\cite{alert2020physical, blanch2018turbulent}, where the velocity correlation lengths  may reach values hundreds of times larger than the typical cell size~\cite{petitjean2010velocity,henkes2020dense,garcia2015physics}.

Current explanations of these collective phenomena are based on macroscopic hydrodynamic-like theories~\cite{wensink2012meso, dunkel2013fluid, james2018turbulence}, in the spirit of the Toner-Tu approach, or, at the microscopic level, by invoking effective alignment interactions in the dynamics of both cell monolayers~\cite{sepulveda2013collective, smeets2016emergent, alert2020physical, sarkar2020minimal} and bacteria~\cite{grossmann2014vortex}.
However, recently, the occurrence of finite size domains where the particles are aligned and, in general, spatial velocity correlations are detected, have been observed also in active dynamics of spherical particles in the absence of explicit alignment interactions, both in phase-separated~\cite{caprini2020spontaneous} and homogeneous configurations~\cite{caprini2020hidden,caprini2020time}. In particular, the models usually employed to describe the behavior of spherical active particles - that are based on independent active forces with a certain degree of persistence and pure repulsive interactions - are enough to reproduce the salient features of the spatial patterns of the velocity field. 
In the infinite volume limit, the theory of Refs.~\cite{caprini2020spontaneous,caprini2020hidden} is able to predict an exponential-like shape of the spatial velocity correlations with a correlation length that increases with the persistence time~\cite{caprini2020spontaneous} and is reduced by inertial forces~\cite{caprini2020spatial} and in the low-density regime~\cite{caprini2020hidden}.
The theory, originally developed for active solid configurations, assuming the 6-fold symmetry, has been recently extended to active liquids~\cite{szamel2021long} where the correlation length of the longitudinal modes is larger with respect to that of the transversal modes.

On the experimental side, active cell monolayers, bacteria, and self-propelled colloidal Quincke rollers show a rich behavior when the system is confined in circular geometries~\cite{bricard2013emergence,zhang2020oscillatory,jain2020role,liu2021viscoelastic}.
Indeed, these systems display a transition from an isotropic phase where the velocities of each particle are slightly correlated to another polar phase where the whole system rotates persistently in the clockwise or anticlockwise direction, even forming a giant vortex spanning the whole system size.
Specifically, self-propelling Quincke rollers in narrow  closed channels may also display coexistence between a polar liquid and a moving solid front of particles~\cite{geyer2019freezing}.
Collective rotations were also observed and studied in highly packed three-dimensional grains enclosed in a cylindrical container~\cite{scalliet2015cages, plati2019dynamical, plati2021non}, where each granular particle is activated by the vibration of the bottom plate.
Finally, water droplets confined in narrow channels, which self-propel due to the Marangoni effect, are able to display a collective motion moving as trains of particles~\cite{Dauchot2020Speed}.


The broad experimental interest in strongly confined active particles motivates the present paper, where we numerically study a system of two-dimensional active disks confined by a narrow annulus.
Here, we show that explicit alignment interactions between the particle velocities of active forces are not necessary to induce the collective rotations, in analogy with the spatial correlations observed in unconfined systems.
The global alignment of the tangential velocity field spontaneously arises from the interplay between repulsive interactions and persistent active forces, in the regime of large persistence time and density. 
In the opposite regime, particles do not align along with the whole system but form multiple domains with correlated velocities. 
The paper is structured as follows: in Sec.~\ref{Sec:Model}, we introduce the model employed to describe active particles confined on a ring-like geometry, while, in Sec.~\ref{sec:3}, we report the main numerical and analytical results to explain and characterize both the stuck and the rotating phase. In particular, we study the polarization of the velocity field for different values of the persistence time and system size.  
This study is corroborated by the numerical and analytical investigation of the spatial velocity correlations (and their correlation length) in both regimes.
Finally, a summary of results and discussion concerning the possible applications of this study is presented in the conclusions.

\begin{figure*}[!t]
\centering
\includegraphics[width=0.85\linewidth,keepaspectratio]{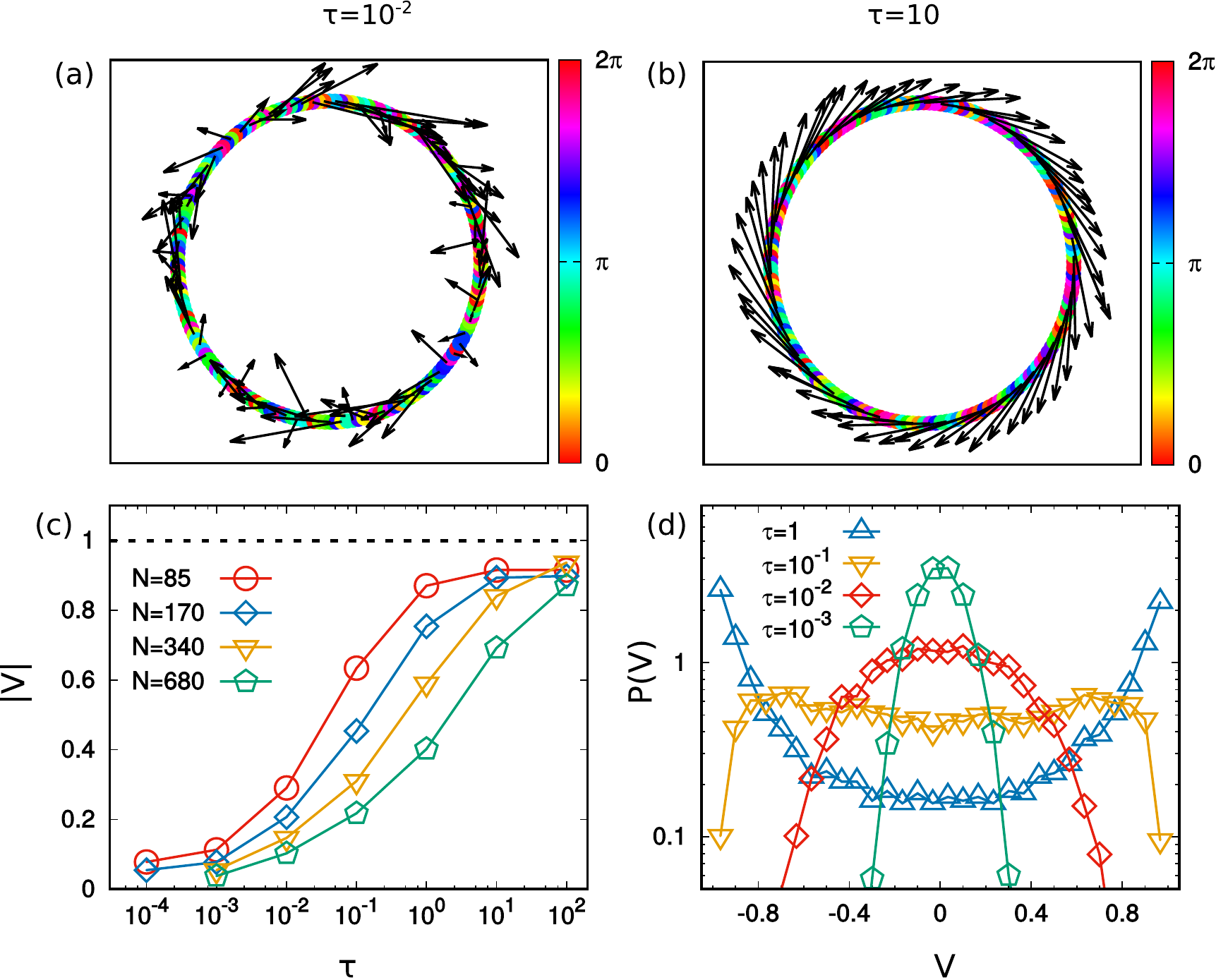}
\caption{\label{fig:SnapMag}
Global velocity alignment. Panels (a) and (b): snapshot configurations on the plane $x,y$ obtained for two different values of $\tau=10^{-2}, 10$, respectively. The colors represent the orientation of the particle active force, $\theta_i$, while the black arrows are the particle velocity (for presentation reasons, we draw 1/3 of the velocity vectors).
Panel (c): modulus of the velocity polarization, $\langle|V|\rangle$, as a function of $\tau$, for different size of the systems, $N=L/\bar{r}$, as indicated in the legend. Panel (d): probability distribution of the polarization, $P(V)$ for different values of $\tau$ as reported in the legend, for $N=170$.
The other parameters are $T=10^{-1}$, $\gamma=10^2$, $v_0=50$ and $\epsilon=10^2$.
}
\end{figure*}

\section{Model}
\label{Sec:Model}

We study a two-dimensional system of $N$ interacting active particles using the underdamped version of the Active Brownian Particles (ABP) model~\cite{takatori2017inertial,scholz2018inertial,mandal2019motility,um2019langevin,lowen2020inertial,sprenger2021time,gutierrez2020inertial,caprini2021inertial}.
The particles are spatially confined in an annular container created by the presence of repulsive soft walls which will be described in the following. 
The particle positions, $\mathbf{x}_i$, and velocities, $\mathbf{v}_i$ evolve according to the law:
\begin{subequations}
\label{eq:motion}
\begin{align}
\dot{\mathbf{x}}_i&=\mathbf{v}_i \,,\\
m\dot{\mathbf{v}}_i& = - \gamma \mathbf{v}_i +\mathbf{F}_i + \mathbf{F}_i^w + \mathbf{f}^a_i + \sqrt{2 \gamma  T} \,\boldsymbol{\eta}_i \,,
\end{align}
\end{subequations}
where the constant $\gamma$ is the drag coefficient, $m$ the particle mass, and $T$ the solvent temperature, that is related to the translational diffusion coefficient, $D_t$, through the Einstein relation, $\gamma D_t= T/m$.
The term $\boldsymbol{\eta}_i$ is a white noise vector with zero average and unit variance accounting for the collisions between the solvent and the active particles, such that $\langle \boldsymbol{\eta}_i(t) \boldsymbol{\eta}_j(t')\rangle=\boldsymbol{\delta}(t-t')\delta_{ij}$.
In analogy with equilibrium colloids, the solvent exerts a Stokes drag force proportional to the particle velocity. 
The particle interactions are represented by the force $\mathbf{F}_i = - \nabla_i U_{tot}$, where $U_{tot} = \sum_{i<j} U(|\mathbf{x}_i -\mathbf{x}_j|)$ is a pairwise potential. 
The shape $U$ is chosen as a shifted and truncated Lennard-Jones potential:
\begin{equation}
\label{eq:interactionpotential}
U(r) = 4\epsilon \left[ \left(\frac{\sigma}{r}\right)^{12}- \left(\frac{\sigma}{r}\right)^6  \right] \,,
\end{equation}
for $r\leq 2^{1/6}\sigma$ and zero otherwise.
The constants $\epsilon$ and $\sigma$ determine the energy unit and the nominal particle diameter, respectively.
The term $\mathbf{F}_i^w$ represents the force exerted by the walls of the container. This is an annulus centered at the origin with inner radius, $R_{in}$, and outer radius, $R_{out}$, so that the average radius is $\bar{R}=(R_{out}+R_{in})$ and its width $w=(R_{out}-R_{in})$.
The repulsion exerted by each wall is modeled through the same potential $U(r)$, introduced in Eq.~\eqref{eq:interactionpotential}, pointing in the radial direction both for the inner and outer walls. 
Further details about the implementation of the wall potentials are reported in Appendix \ref{app:details}.

In the ABP model, the active force is chosen as a time-dependent force, $\mathbf{f}_i^a$, with a stochastic evolution, that acts locally on each particle~\cite{fily2012athermal,buttinoni2013dynamical,mognetti2013living,solon2015pressure,digregorio2018full,hecht2021introduction}.
At this level of description, the details about the microscopic-system-dependent origin of $\mathbf{f}_i^a$ are not specified.
This force drives the system out of equilibrium and determines a persistent motion in a random direction lasting for a time smaller than a characteristic  persistence time, $\tau$.
According to the ABP model, the active force, $\mathbf{f}^a_i=f_0 \mathbf{n}_i$, has constant modulus $f_0$ and a time-dependent orientation, $\mathbf{n}_i=(\cos{\theta_i}, \sin{\theta_i})$. The angle $\theta_i$ evolves stochastically via a Brownian motion:
\begin{equation}
\label{eq:theta}
\dot{\theta}_i = \sqrt{2D_r} \chi_i \,,
\end{equation}
where $\chi_i$ is a white noise with zero average and unit variance and $D_r=1/\tau$ determines the persistence time of the active force. We also remark that $f_0$ fixes the swim velocity induced by the self-propulsion, namely $v_0=\frac{f_0}{\gamma}$, which is smaller as $\gamma$ is increased.
The value of $D_t$ is chosen smaller than the effective diffusivity due to the active force, $D_a=v^2_0 \tau$, as in typical experimental conditions of active colloidal and bacterial suspensions~\cite{bechinger2016active}.
We also fix $\gamma=10^2$ and $m=1$ so that the inertial time reads $\tau_I=m/\gamma=10^{-2}$.

\section{Results}\label{sec:3}

\begin{figure*}[!t]
\centering
\includegraphics[width=0.9\linewidth,keepaspectratio]{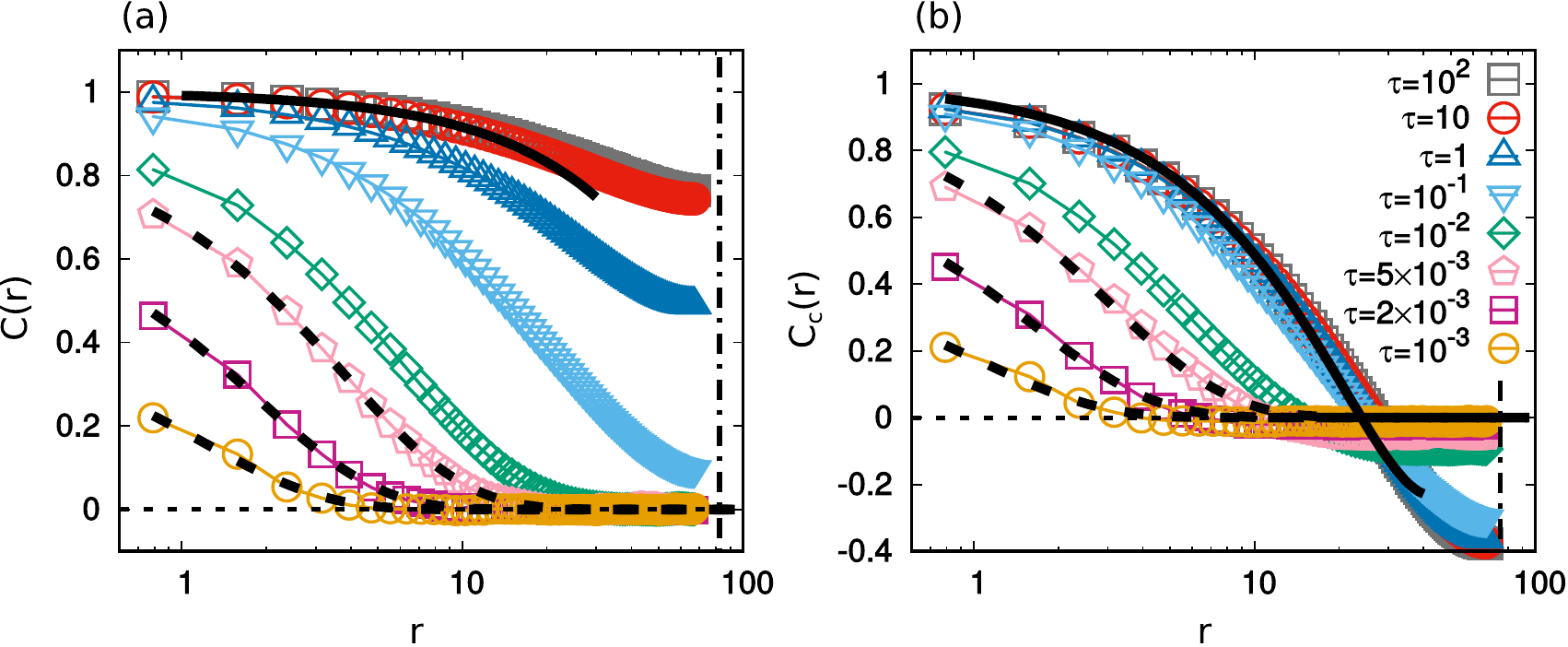}
\caption{\label{fig:Correlations}
Spatial velocity correlations. Spatial correlations of the tangential velocity, $C(r)$, and its connected version, $C_c(r)$, for panel (a) and (b), respectively.
The correlations are reported  for different values of $\tau$ as shown in the legend of panel (b) which is shared also by panel (a).
The solid black curve in panel (a) plots the phenomenological curve $1-r/\bar{r}/N$, while the solid black line in panel (b) corresponds to the theoretical prediction~\eqref{eq:Cc(r)_prediction}. 
The dashed black lines in both panels are obtained from the other theoretical prediction~\eqref{eq:lambda_prediction_small}.
Finally, the other black lines are eye-guides, the dashed horizontal one marks zero while the dot-dashed vertical one determines the maximal distance that the system is allowed to explore corresponding to $L=\pi \bar{R}$.
The other parameters are $T=10^{-1}$, $\gamma=10^2$, $v_0=50$, $N=170$ and $\epsilon=10^2$.
}
\end{figure*}

The dynamics~\eqref{eq:motion} has been numerically integrated setting the channel width so that it equals the particle diameter.
In this way, the system can be treated as an effective one-dimensional system with packing fraction, $\phi \approx N/L \bar{r}$, where $L=2\pi \bar{R}$ is the length of the ring and $\bar{r}$ the average distance between neighboring particles along the ring. 
In the numerical study, $\phi$ is kept constant and it is chosen to be large enough so that the system displays almost solid-like one-dimensional configurations. In this regime, the $i$-th particle interacts with the $(i+1)$-th and $(i-1)$-th particle which are separated by an average distance $\bar{r}=\pi \bar{R}/N$ from the $i$-th particle.
Such a ``chain'' of active particles is positioned at distance $\bar{R}=(R_{in}+R_{out})/2$ from the center of the circular crown.

Fig.~\ref{fig:SnapMag}~(a) and~(b) shows two different snapshot configurations obtained for $\tau=10^{-2}, 10$, respectively. The color gradients are chosen according to the direction of the active force confirming that they are randomly distributed.
Each black arrow draws the particle velocity and reveals a fascinating scenario.
For the larger $\tau$ value, the particle velocities are aligned along the tangential direction forming a unique domain spanning the whole ring. In this regime, the particles move coherently along the tangential direction revealing a clockwise (or an anti-clockwise) motion even though there are no explicit forces responsible for such a global alignment. 
This phenomenon disappears for smaller values of $\tau$ where one can still observe the formation of small domains where the tangential component of the particle velocities are aligned even if the system does not show collective rotations.
Further decrease of $\tau$ (corresponding to further increase of $D_r$) allows the active system to behave as a passive one without spatial velocity correlations and with effective temperature $\sim v_0^2 \gamma/D_r$, a limit that in the absence of inertia has been investigated both for interacting and non-interacting active systems~\cite{caprini2019transport} (See also Refs.~\cite{cugliandolo2019effective,petrelli2020effective} for recent works on the effective temperature in active systems). 
In what follows, we identify the large persistence regimes as those values of $\tau$ showing collective rotations while we call small persistence regimes the remaining smaller values of $\tau$.

\subsection{Polarization of tangential velocity}

To give a quantitative measure of the alignment degree characterizing the spontaneous rotations occurring in the system, we introduce the instantaneous collective polarization of the velocity, $V(t)$, defined as:
\begin{equation}
\label{eq:polarization}
V(t)= \frac{1}{N}\sum_i \frac{ v^t_i(t) }{|v^t_i(t)| } \,,
\end{equation}
where $v^t_i$ is the tangential component of the velocity with respect to the center of the annulus.
The variable $V(t)$ is not to be confused with the polarization of the active force that is trivially zero in this system because the self-propulsions evolving through Eq.~\eqref{eq:theta} are independent of each other.
$V(t)$ has the following properties: i) its temporal average vanishes, $\langle V\rangle=0$, since  no forces break the rotational symmetry for finite $\tau$ ii) $V(t)$ reads almost zero if $v^t_i(t)$ are independent of each other while takes the values $1$ and $-1$ for clockwise or anticlockwise rotating configurations, respectively, occurring when the particle velocities are globally aligned.
Similarly to Ising-like models, $V(t)$ can be interpreted as a sort of magnetization and, thus, a useful way to take the temporal average without losing the information about the alignment is to consider the average of its absolute value, $\langle |V| \rangle$.
This observable is shown in Fig.~\ref{fig:SnapMag}~(c) as a function of $\tau$, at fixed active speed, $v_0$, and for different system size $L=\bar{r}N$ and confirms the qualitative scenario already observed qualitatively in the snapshot configurations.
In particular, for the whole range of system size explored, $\langle |V|\rangle$ monotonically increases with $\tau$, from a very small value, that is $\sim 0$, until to a large value $\sim 0.9$ where the particles are well-aligned to each other and the system shows spontaneous collective rotations.
This scenario is confirmed by the study of the distribution of $V(t)$ reported in Fig.~\ref{fig:SnapMag}~(d):
for the smaller values of $\tau$ of the graph, $\mathcal{P}(V)$ has a Gaussian-like profile  peaked around the origin and, in this regime, increasing $\tau$ simply broadens the distribution.
In a further regime of $\tau$, the distribution develops pronounced deviation from the Gaussian shape and, in particular, at some {\it threshold} value, two symmetric peaks are formed and the distribution becomes bimodal.
In this regime, the increase of $\tau$, on the one hand, shifts the peaks towards $1$ and $-1$ and, on the other hand, produces higher and narrow peaks with a consequent very small probability of having $V(t) \sim 0$.
However, as expected by symmetry, clockwise and anti-clockwise rotations occur with the same probability (even in the presence of collective rotations) as confirmed by the shape of $\mathcal{P}(V)$.

The system size does not change qualitatively the picture so far presented and, in particular, the monotonic increase of $\langle |V| \rangle$ with $\tau$.
However, the larger $L=\bar{r}N$, the smaller $\langle |V| \rangle$ (at fixed $\tau$), so that the occurrence of spontaneous rotations needs larger values of $\tau$ for increasing $L$. 
This is a first clue that the scenario presented (and, in particular, the collective phase) does not survive the infinite volume limit and, thus, does not correspond to a phase-transition in the usual thermodynamic sense.
Moreover, it is still remarkable that the finite size of the system can induce a transition from a disordered state, not showing global rotations, to an ordered state, characterized by collective rotations, despite the absence of explicit alignment interactions that couple particle velocities or active forces.

\subsection{Spatial velocity correlations and correlation length}

\begin{figure*}[!t]
\centering
\includegraphics[width=0.9\linewidth,keepaspectratio]{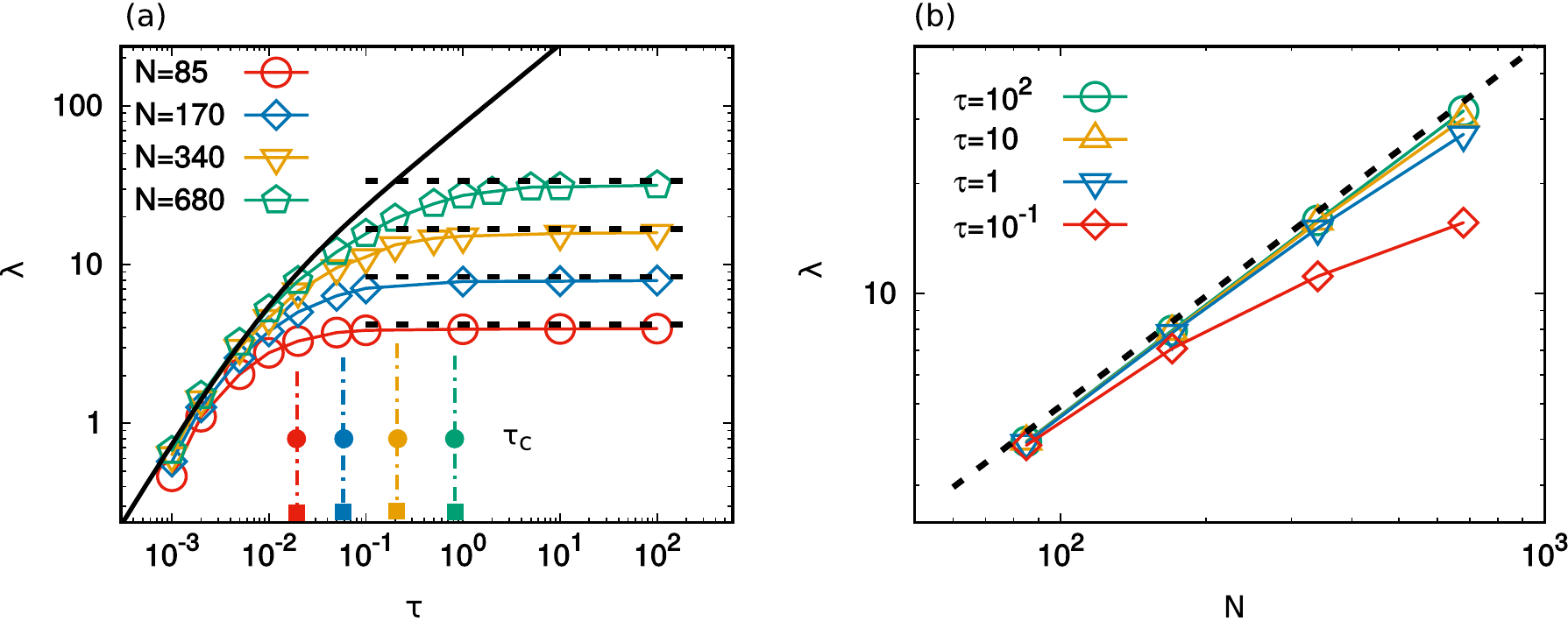}
\caption{\label{fig:lambdar0}
Correlation length. Panel~(a): correlation length, $\lambda$, as a function of $\tau$, for different system size $N=L/\bar{r}$ as indicated in the legend. The dashed black horizontal line are obtained from Eq.~\eqref{eq:scalefree_lambda} for each system size, while the solid black line plots Eq.~\eqref{eq:lambda_prediction_small}.
The vertical dotted dashed lines are marked in correspondence of  $\tau=\tau_c$ for each system size according to the color notation of the legend. The $\tau_c$ values are obtained from Eq.~\eqref{eq:tauc_prediction}.
Panel (b): $\lambda$ as a function of $N$ for different values of $\tau$ as indicated in the legend. Here, the black dashed line is obtained from Eq.~\eqref{eq:scalefree_lambda}.
The remaining parameters are $T=10^{-1}$, $\gamma=10^2$, $v_0=50$ and $\epsilon=10^2$.
}
\end{figure*}

Disordered and ordered states are studied in terms of the spatial connected correlation function, $C_c(r)$~\cite{cavagna2020equilibrium,cavagna2021novel,cavagna2018physics}, that provides the information about the spatial correlation between observables at separation $r$.
To capture the effective one-dimensional aspect of the system, we study the spatial correlation of the velocity component tangent to the ring.
We first introduce the correlation, $C(r)$ as:
\begin{equation}
C(r)=\frac{\langle v^t(r) v^t(0) \rangle}{\langle (v^t)^2 \rangle} \,,
\end{equation}
normalized with respect to the second moment of the tangential velocity, $\langle (v^t)^2 \rangle$, and the connected correlation defined as:
\begin{equation}
C_c(r)=\frac{\langle \delta v^t(r)  \,\delta v^t(0) \rangle}{\langle [\delta v^t(0)]^2 \rangle} \,,
\end{equation}
where $\delta v^t(r)= v^t(r) - \bar{v}^t$ represents the deviation of the velocity variable from its spatial average $\bar{v}^t=1/N\sum_i v^t_i$ and $r$ is the spatial coordinate along the ring.
The argument of both correlations cannot exceed the maximal distance along with the ring, $\sim \pi \bar{R}$.
We remark that using the connected velocity correlation function one can define the correlation length even in the case of non-ergodic systems and, in particular, when the spatial average, $V(t)$, over the whole system does not vanish~\cite{cavagna2020equilibrium} (as found for the larger values of $\tau$). To include these possibilities, the correlation length $\lambda$ is defined as~\cite{cavagna2020equilibrium}:
\begin{equation}
\label{eq:def_lambda}
\lambda  = \frac{\int_0^{r_0} r \, C_c(r) \, dr}{\int_0^{r_0} C_c(r) \, dr} \,.
\end{equation}
where $r_0$ is the distance where $C_c(r_0)=0$. In the case of an exponential decay  where $C_c(r_0)\geq 0$, we assume $r_0\approx \pi \bar{R}$ (see Ref.~\cite{cavagna2018physics} for a recent review on such a method).

\subsubsection{Small persistence regime}

$C(r)$ and $C_c(r)$ are shown for several values of $\tau$ in Fig.~\ref{fig:Correlations}~(a) and~(b), respectively.
In the small $\tau$ regime  (i.e. when the system does not display global collective rotations, $\langle|V|\rangle \approx 0$), $C(r)=C_c(r)$. 
Both observables  decay exponentially with a typical correlation length that increases with $\tau$.
The Fourier transform of the
tangential velocity correlation has the following form (see Appendix~\ref{appendix:Fourier}):
\begin{equation}
\label{eq:G0}
\langle \hat{v}^t(q)\hat{v}^t(-q)\rangle \propto \frac{1}{1+2 \frac{\ell^2}{\bar{r}^2} [1-\cos(q)]} \,,
\end{equation}
where $q=  2\pi j/N$, with $j=0, 1, 2, ..., N-1$, is a one-dimensional wave-vector belonging to the reciprocal Fourier space and $\hat{v}^t(q)$ is the Fourier transform of the tangential velocity. We have also assumed $T\ll v_0^2$ as observed in the experiments (the full expression is reported in Appendix~\ref{appendix:Fourier}).
The length scale, $\ell$, can be expressed in terms of the model parameters:   
\begin{equation}
\label{eq:lambda_prediction_small}
\ell^2  =   \bar{r}^2 \frac{U''(\bar{r})}{m} \frac{\tau^2}{1+\frac{\tau}{\tau_I}}  \,.
\end{equation}
and depends both on the density (via the second derivative of the potential and $\bar{r}$) and on the typical relaxation times governing the dynamics, namely $m/\gamma$ and $\tau$.
The $q$-space correlation~\eqref{eq:G0} can be  transformed back to real space in the limit $\ell \ll L$, as shown in Appendix~\ref{appendix:smallpersistence}, leading to an exponential behavior in agreement with Fig.~\ref{fig:Correlations}~(a) and~(b):
\begin{equation}
\label{eq:exp_shap_C}
C_c(r)=C(r)\propto e^{-r/\ell} \,.
\end{equation}
The prediction~\eqref{eq:exp_shap_C} is in good agreement with the numerical data as revealed in Fig.~\ref{fig:Correlations}~(a) and~(b) for the smaller values of $\tau$ reported in the numerical study (see the comparison between colored data and dashed black lines). 
This range depends on the system size and is larger as $L=\bar{r}N$ is increased. We also remark that, when the prediction~\eqref{eq:exp_shap_C} holds, the dynamical parameter $\ell$ coincides with the correlation length, $\lambda$, as can be seen from its definition~\eqref{eq:def_lambda}.
This agreement is numerically confirmed in Fig.~\ref{fig:lambdar0}~(a) comparing $\lambda$ (colored points) and the prediction~\eqref{eq:lambda_prediction_small} (solid black lines) for different values of the system size.
The agreement between data and theory holds up to a threshold value, $\tau^*$ that increases when $L=\bar{r}N$ is increased as expected from the analysis of the previous section.
In particular, in this regime such that $\lambda\approx \ell$, $\lambda$ scales as $\tau/\sqrt{1+\tau/\tau_I}$ so that $\lambda \propto \tau^{1/2}$ in the overdamped limit when $\tau_I \ll \tau$ and $\lambda \propto \tau$ in the opposite inertial regime such that $\tau_I \gg \tau$. 
Moreover, in both cases, $\lambda$ is not affected by the system size.
This correlation length determines the average size of the domains where the velocities are correlated and, thus, aligned. 
Besides, the condition $\lambda=\ell \ll L=\pi \bar{R}$, holding in this regime, guarantees the presence of many domains along with the whole ring with different velocity directions, in such a way that $V(t)\approx 0$.We also observe that if $\lambda=\ell \lesssim \bar{r}$ the spatial velocity correlation are negligible (being smaller than the particle diameter) and the system behaves as a passive system with almost uncorrelated particle velocities.

As a further remark, the correlation length $\lambda$ depends only on $\tau$ and the inertial time, $m/\gamma$. In agreement with previous theoretical results on two-dimensional infinite systems~\cite{caprini2020spatial}, $\lambda$ does depend neither on the swim velocity, $v_0$ nor on the solvent temperature $T$.
In other words, the dynamical phenomenon reported here has not thermal origin and is a dynamical collective effect.

\subsubsection{Large persistence regime}

For the larger values of $\tau$ such that $V(t) \neq 0$, the system is non-ergodic and the spatial velocity correlations, shown in Fig.~\ref{fig:Correlations}, reveal an interesting behavior.
For $\tau \gtrsim \tau^*$, the decay becomes slower than exponential and, in particular, $C(r)$ does not decay towards zero, as emerged by Fig.~\ref{fig:Correlations}~(a).  
This is because the system displays a non-zero polarization of the velocity, as previously discussed, and 
confirms that when the spontaneous rotations take place the particle velocities of the whole systems are strongly correlated.
When $C(r)$ does not decay to zero, $C_c(r)$ starts differing from $C(r)$.
The profiles of $C_c(r)$ for different values of $\tau$ are reported in Fig.~\ref{fig:Correlations}~(b) for a given value of the system size taken as a reference case.
In particular $C_c(r)$ goes below zero at some value $r_0$ which varies as $\tau$ is increased until a saturation occurs.
In this case, the profiles of $C_c(r)$ for large values of $\tau$ collapse onto the same curve (roughly for $\tau \gtrsim 10^{-1}$). This saturation profile displays an algebraic decay that results in good agreement with the theoretical prediction derived in Appendix~\ref{appendix:largetau}, which reads:
\begin{equation}
\label{eq:Cc(r)_prediction}
C_c(r) \approx 1 - \frac{r}{\bar{r}}\frac{\pi^2}{N} +  \frac{r^2}{\bar{r}^2} \frac{2\pi^2}{N^2} \,.
\end{equation}
This expression holds up to $O\left(r^4/(N^4\bar{r}^4) \right)$ and, thus, becomes inaccurate as $r/N$ increases but shows a good agreement at least up to $r\approx r_0$ (the value such that $C_c(r_0)=0$).  For this reason, it can be employed in the calculation of $\lambda$.
The condition to get Eq.~\eqref{eq:Cc(r)_prediction} is that $\ell \gg L$, a parameter that, in this regime, does not represent anymore the correlation length of the system.
It is remarkable that, according to Eq.~\eqref{eq:Cc(r)_prediction}, $C_c(r)$ only depends on the system size, $L=\bar{r}N$ and on the distance between neighboring particles, $\bar{r}$. The other parameters, such as persistence time, swim velocity, viscosity, and temperature, are completely irrelevant in this regime.

Fig.~\ref{fig:lambdar0}~(a) and~(b) shows $\lambda$ as a function of $\tau$ for different system size, $L=\bar{r} N$, and as a function of $L=\bar{r} N$ for different $\tau$, respectively.
As already observed, $\lambda$ increases with $\tau$ and does not depend on the system size for the smaller values of $\tau$ in agreement with the theoretical prediction~\eqref{eq:lambda_prediction_small}. This holds up to a threshold value, $\tau^*$, that increases with $L$.
A dependence on $L=\bar{r}N$ emerges for $\tau>\tau^*$ (panel~(a)) that lowers $\lambda$ with respect to the value predicted by Eq.~\eqref{eq:lambda_prediction_small}. 
The smaller $L$, the larger the discrepancy with this prediction. 
In practice, the size of the system acts as a natural cut-off for the correlation length.
At some value of $\tau$, namely $\tau_c$, that is again determined by the system size, the value of $\lambda$ saturates to $\lambda_c$ that is $\tau$ independent and scales linearly with $L$, as clearly shown in panel~(b). 
The dependence on $L$ becomes slower than a linear function when $\tau<\tau_c$.
The value of $\lambda_c$ can be theoretically predicted by its definition~\eqref{eq:def_lambda}, using the approximated profile of $C_c(r)$, namely Eq.~\eqref{eq:Cc(r)_prediction}.
Indeed, it is possible to analytically calculate, the value of $r_0$ such that $C_c(r_0)=0$, that reads:
\begin{equation}
\label{eq:r0_vs_tau}
r_0\approx \alpha L \,,
\end{equation}
where $\alpha=\frac{1}{4} \left( 1-\sqrt{1-\frac{8}{\pi^2}}  \right) < 1$ is a numerical factor that does not depend on the parameters of the model and on the system size.
Plugging the prediction~\eqref{eq:r0_vs_tau} into the definition of $\lambda$ (Eq.~\eqref{eq:def_lambda}) and using the explicit expression for $C_c(r)$, we obtain:
\begin{equation}
\label{eq:scalefree_lambda}
\lambda \approx \beta L \,,
\end{equation}
where $\beta$ is another numerical constant, depending neither on system size nor on the parameters of the active force, which reads: 
\begin{equation}
\beta=\alpha \left( \frac{1}{2} - \frac{1}{12} \pi^2 \alpha +\frac{\pi^2}{6}\left(1-\frac{\pi^2}{4}\right)\alpha^2 \right) + O(\alpha^3) \,.
\end{equation}
The prediction~\eqref{eq:scalefree_lambda} is in fair agreement with the numerical data as shown both in Fig.~\ref{fig:lambdar0}~(a) and~(b), confirming that, in this regimes of parameters, $\lambda$ (as also $C_c(r)$) does not depend on the parameters of the model but is purely determined by the size of the system.

\subsection{Absence of criticality in the infinite volume limit}

Fig.~\ref{fig:lambdar0} (a) and (b) indicate the absence of any criticality or scale-free properties surviving to the infinite volume limit. 
This finding can be rationalized by remarking that, after expanding Eq.~\eqref{eq:G0} for small $q$,
the Fourier transform of the tangential velocity correlation has the same Ornstein-Zernike form as the mean-field spin-spin correlation
 of the one-dimensional Ising model.
However, at variance with the Ising model, since $\ell^2>0$ there are no values of the parameters for which Eq.~\eqref{eq:G0} diverges in the infinite volume limit, i.e. for $q$ values arbitrarily small.
In other words, a system of ABP particles does not show any criticality in the infinite volume limit for finite $\tau$. 
In our periodic geometry, this limit can be achieved by setting $\bar{R}\to \infty$, but, in practice, coincides with the condition $\ell \ll L$, that leads to the exponential prediction~\eqref{eq:exp_shap_C}.

Here, we focus attention on finite-size periodic systems (similar to those experimentally analyzed in Refs.~\cite{zhang2020oscillatory,jain2020role,liu2021viscoelastic}), where the regime $\ell \gg L$ is  accessible even experimentally.
In this case, the expression for $\langle \hat{v}(q)\hat{v}(-q)\rangle$ is dominated by the contribution for small $q$. 
Subtracting the term corresponding to the zero mode, $q=0$, from $\langle \hat{v}(q)\hat{v}(-q)\rangle$ in Eq.~\eqref{eq:G0}, and taking into account the finite size of the system, the first accessible value of $q$ is $q_{min}=2\pi/N$. 
By defining $\delta \hat{v}(q)=\hat{v}(q)-\hat{v}(0)$, if the following condition holds
\begin{equation}
\label{eq:condition_tauc}
\ell^2\gg 1/q_{min}^2=\frac{L^2}{4\pi^2} = \ell^2_c \,,
\end{equation}
we can approximate
\begin{equation}
\label{eq:vqvq}
\langle \delta\hat{v}(q)\delta\hat{v}(-q)\rangle \propto \frac{1}{\ell^2[1-\cos(q)]} \,,
\end{equation}
with $q=(2\pi/N) j$ with $j=1, 2, ... , N-1$ (we remind that we have assumed $T \ll v_0^2$ to get Eq.~\eqref{eq:vqvq}).
We also remark that, upon normalizing Eq.~\eqref{eq:vqvq}, the profile of the normalized spatial velocity correlation does not depend on the details of the model but just on the system size as observed in the numerical study for the larger values of $\tau$. 
Therefore, the condition~\eqref{eq:condition_tauc} allows us to define a  ``crossover'' value $\tau_c$ from $\ell_c$, below which the prediction~\eqref{eq:scalefree_lambda} fails:
\begin{equation}
\label{eq:tauc_prediction}
\begin{aligned}
\tau_c = \frac{L^2}{  8\pi^2} \frac{\gamma}{\bar{r}^2 U''(\bar{r})} \left(1+\sqrt{1+\frac{16\pi^2}{L^2} \bar{r}^2 U''(\bar{r}) \frac{m}{\gamma^2}   }\right) \,.
\end{aligned}
\end{equation}
The predictions from~\eqref{eq:tauc_prediction} are plotted in Fig.~\ref{fig:lambdar0}~(a) as vertical dashed dotted lines for each system size, $L=\bar{r}N$. This analysis confirms that Eq.~\eqref{eq:tauc_prediction} is a good marker to select the range of $\tau$ values such that $\lambda$ reaches its plateau.
We also remark that the value of $\tau_c$ increases as $L^2$ when subleading orders in powers of $L$ are neglected. This is a further confirmation that the predictions~\eqref{eq:Cc(r)_prediction} and~\eqref{eq:scalefree_lambda} cannot hold in the infinite volume limit.

\subsubsection{The finite-size scaling ansatz}

\begin{figure}[!t]
\centering
\includegraphics[width=0.9\linewidth,keepaspectratio]{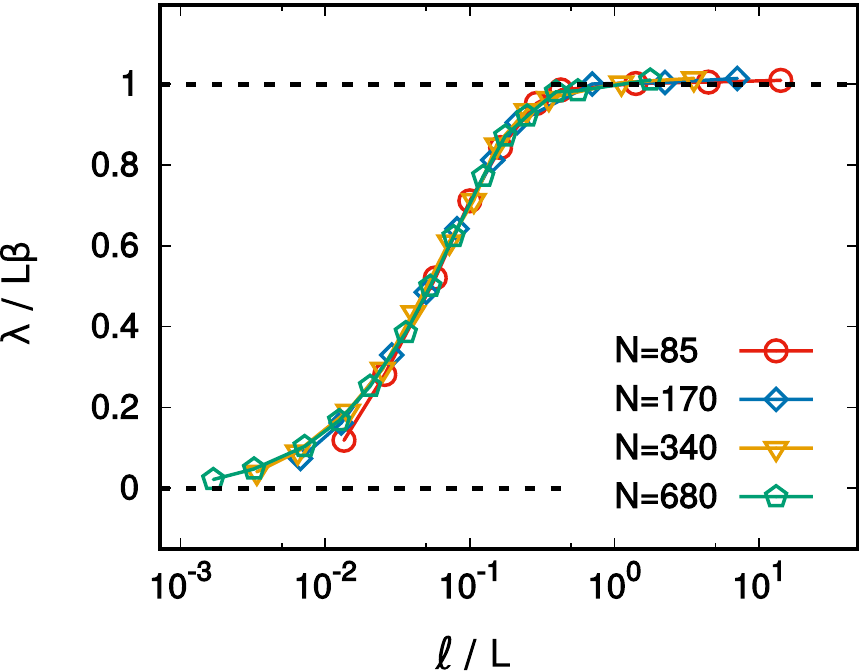}
\caption{\label{fig:scaling}
Rescaled correlation length, $\lambda/(L\beta)$, as a function of $\ell/L$, for different system size $N=L/\bar{r}$ as indicated in the legend. The dashed black horizontal lines are eye guides marked in correspondence of zero and $\lambda/(L\beta)=1$.
The other parameters are $T=10^{-1}$, $\gamma=10^2$, $v_0=50$ and $\epsilon=10^2$.
}
\end{figure}

The spatial velocity correlation function discussed so far is characterized by two length scales that are the infinite-system correlation length $\ell$ (given by Eq.~(\ref{eq:lambda_prediction_small})) and the size of the system $L$.
$\ell$ coincides with the correlation length of the system, $\lambda$, in the small persistence regime, where $\ell \ll L$.
On the other hand, $L$ is the only relevant length scale in the large persistence regime, where $\ell \gg L$, and $\lambda$ becomes independent on the model parameters being only determined by $L$ according to Eq.~\eqref{eq:scalefree_lambda}.
Unfortunately, in the regime of parameters such that $\ell \approx L$ (roughly corresponding to $\tau^*\lesssim\tau\lesssim\tau_c$) $\lambda$ cannot be easily predicted theoretically. However, by following standard scaling arguments~\cite{fisher1972scaling,binder2012monte}, we expect that a smooth function of the ratio $\ell/L$ describes the behavior of $\lambda$ also in this cross-over regime.
To corroborate this hypothesis we formulate the following finite-size scaling ansatz:
\begin{equation}
\label{eq:scaling_law}
\lambda = L \, g(\ell/L) \,,
\end{equation}
where $g(\ell/L)$ is a function whose detailed form is unknown except for its asymptotic behavior, that can be extrapolated by our theoretical arguments: ${g(\ell/L) = \mathrm{const}}$ as $\ell/L\rightarrow \infty$ and $g(\ell/L)=\ell/L$ when $\ell/L\rightarrow 0$. 
Note also that the consistency with Eq.~(\ref{eq:scalefree_lambda}) implies ${g(\ell/L\rightarrow \infty) =\beta}$.
The scaling law~\eqref{eq:scaling_law} is checked in Fig.~\ref{fig:scaling} where $\lambda/(\beta L)$ is plotted as a function of $\ell/L$ for several values of $\tau$ and $L$. Data with different values of $\tau$ but the same $L$ are plotted with the same color, revealing a good data collapse. This confirms the validity of the ansatz~\eqref{eq:scaling_law} in  the whole range of parameters analyzed so far.

\section{Conclusion}
\label{Sec:Conclusion}

In this article, we have studied a system of repulsive active particles evolving with the underdamped ABP model confined to an annular region by soft walls. Despite the absence of explicit alignment interactions between the particle velocities and/or active forces, the particles synchronize showing the occurrence of velocity alignment producing collective rotations.
In particular, when the persistence of the active force increases, our system shows a transition from i) a stuck disordered state  to ii) a globally ordered state characterized by a collective rotating motion that alternates clockwise and anti-clockwise rotations.
The state i) is characterized by an almost vanishing polarization of the velocity and by exponential profiles of the spatial velocity correlations, whose correlation length is independent of the system size and increases with the persistence time in agreement with previous studies.
In the rotating state ii), the velocity polarization reaches large values and the system is non-ergodic since the spatial average of the particle velocities does not coincide with its temporal average.
Moreover, the connected version of the spatial velocity correlations assumes negative values and displays a correlation length that is uniquely determined by the system size and does not depend on the parameters of the active force.
However, these effects (and, in particular, the rotating states) disappear in the infinite volume limit, and thus, they do not  signal a phase transition in the thermodynamic sense. However, it is still remarkable that collective alignment effects spontaneously emerge in finite-size systems confined in a periodic geometry even in the absence of explicit alignment interactions.

It would also be interesting to study how these effective alignments affect the rectification efficiency in asymmetric geometries. It has been indeed observed both in experiments and simulations that the behavior of ratchet motors driven by active particles can be quite erratic~\cite{angelani2009self,di2010bacterial,kaiser2014transport} unless the orientations of the active particles are fixed~\cite{vizsnyiczai2017light,maggi2016self}. It could be that the confinement-induced collective alignment of active particles, as the one studied in this paper, could be exploited to improve the performances of these particles-based micromotors.

\subsection*{Acknowledgements}
LC and UMBM acknowledge support from the MIUR PRIN 2017 project 201798CZLJ
and warmly thank Andrea Puglisi for letting us use the computer facilities of his group and for discussions regarding 
some aspects of this research.


\appendix

\section{Details on the wall geometry}\label{app:details}

In this appendix, we provide the technical details on the wall implementation responsible for the confinement of the particles into an annular region, that is realized through a narrow circular crown.
Both the outer and the inner walls exert a force, $\mathbf{F}^w_{out}$ and $\mathbf{F}^w_{inn}$, respectively, that acts on each particle (in this appendix the particle index is suppressed to simplify the notation).
As stated in Sec.~\ref{Sec:Model}, the two forces are obtained from the truncated and shifted Lennard Jones potential, $U(r)$, described by the profile~\eqref{eq:interactionpotential} (also used to model the repulsion between two active particles) and point radially with the respect to the center of the ring which is placed at the origin.
Specifically, we have:
\begin{flalign}
\mathbf{F}^w_{out} &= -U'(R-\bar{R}_{out}) \hat{R} \\
\mathbf{F}^w_{inn} &= U'(R-\bar{R}_{inn}) \hat{R} \,,
\end{flalign}
where $R$ is the radial coordinate of the particle position (calculated with respect to the center of the ring) and $\hat{R}$ is the unit versor pointing radially (outward with respect to the origin). $R_{out}$ and $R_{inn}$ are the positions of the outer and the inner radius of the circular crown, respectively. 
$U'$ is simply the derivative of the potential $U$ with respect to its argument.
As a consequence, $\mathbf{F}^w_{inn}$ is a force defined for $R>R_{inn}$ while $\mathbf{F}^w_{out}$ for $R<R_{out}$ that confine the radial coordinate of each particle to be in the interval $(R_{inn}, R_{out})$.
Finally, the total force $\mathbf{F}^w$ appearing in the dynamics of each particle (see Eq.~\eqref{eq:motion}) is simply given by 
\begin{equation*}
\mathbf{F}^w=\mathbf{F}^w_{inn}+\mathbf{F}^w_{out} \,.
\end{equation*}
We remark that in the effective one-dimensional system such that $w = R_{out}-R_{inn} \lesssim \sigma$, the force $\mathbf{F}^w$ fixes the radial coordinate to be $R\approx \bar{R}=(R_{out}+R_{inn})/2$ precluding the dynamics on the radial direction.

\section{Radial and tangential coordinates}

Before taking advantage of the circular geometry, it is useful to manipulate the particle interactions, $\mathbf{F}_i$.
Following Refs.~\cite{caprini2020hidden, caprini2020time}, we truncate the interparticle potential, $U_{tot} = \sum_{i<j} U(|\mathbf{x}_i -\mathbf{x}_j|)$ at the first non vanishing order performing a Taylor expansion around the equilibrium interparticle distance.
Our effective one-dimensional geometry allows the particle $i$ to interact only with the particles $i+1$ and $i-1$ (with the exceptions of the particle $i=1$, which interacts with $i=2$ and $i=N$, and of the particle $i=N$, which interacts with  $i=1$ and $i=N-1$, because of the periodicity of the circular geometry).
With these assumptions, $\mathbf{F}_i$ reads:
$$
\mathbf{F}_i\approx- K (2\mathbf{x}_i - \mathbf{x}_{i+1}-\mathbf{x}_{i-1}) \,,
$$
where the constant $K$ is 
$$
K\approx U''(\bar{r})  \,.
$$
The harmonic approximation of the potential works because we are considering systems with a large density such that neighboring particles could just oscillate around their average interparticle distance, $\bar{r}$, by small deviations.

The circular symmetry of the geometry suggests natural coordinates to study the dynamics~\eqref{eq:motion} and develop a suitable theory.
Since the particles are arranged on a ring at distance $\bar{R}$ from the origin, each particle position is described by the
radial coordinate  $R_i$  and the  polar angle $\psi_i$.
The velocity vector of each particle, $\mathbf{v}_i$, could be decomposed into its radial and tangential components $v^r_i$ and $v^t_i$, respectively.
With this choice, we have:
\begin{flalign}
&\dot{R}_i=v^r_i\\
&R\dot{\psi}_i=v^t_i
\end{flalign}
while the components of the velocity  evolve with
\begin{flalign}
&\dot{v}^r_i = \frac{(v^t_i)^2}{R_i} + F^r_i - \gamma v^r_i + (f^a_i)^r + \sqrt{2\gamma T}\eta^r_i + F_i^w\\
\label{eq:tangential_dynamics}
&\dot{v}^t_i = - \frac{v^r_i v^r_i}{R_i}+F^t_i - \gamma v^t_i + (f^a_i)^t + \sqrt{2\gamma T}\eta^t_i \,.
\end{flalign}
In these equations, $\eta_r$ and $\eta_t$ are two white noises with zero average and unit variance, while $F^r_i$ and $F^t_i$ are the radial and tangential components of the force due to the interparticle interactions. 
The same notation applies  to the active force components.
$F^w_i$ is the force due to the walls, constraining the particles on the ring, and acts along with the radial component only.

When the motion is constrained to an annular region (i.e. a ring) we can assume that $\dot{v}^r_i=0$, $v^r_i=0$ and $R_i=\bar{R}$ so that
the dynamics is ruled only by Eq.~\eqref{eq:tangential_dynamics} that further simplies and reads:
\begin{equation}
\label{eq:tangential_dynamics}
\dot{v}^t_i = F^t_i - \gamma v^t_i + (f^a_i)^t + \sqrt{2\gamma T}\eta^t_i \,.
\end{equation}
The tangential component of the force due to the repulsion of the other particles, $\mathbf{F}_i$, can be expressed as:
\begin{equation*}
F^t =- \hat{y} F_x + \hat{x} F_y \,,
\end{equation*}
where $\hat{x}=\cos{\psi}$ and $\hat{y}=\sin{\psi}$ are unit vectors along the $x$ and $y$ directions.
Specifically, the tangential component reads:
\begin{equation}
\label{eq:force_tangential}
\begin{aligned}
F^t_i &\approx -\bar{R} K \left[ \sin{(\psi_i -  \psi_{i+1})} + \sin{(\psi_i -  \psi_{i-1})} \right] \\
&\approx - \bar{R} K \left[2\psi_i - \psi_{i+1}-\psi_{i-1} \right] \\
&=-  K \left[2r_i - r_{i+1}-r_{i-1} \right] \,.
\end{aligned}
\end{equation}
where $r_i=\bar{R} \psi_i$ defines the tangential coordinate along the ring of the $i$-th particle.

The expansion of the sinus function for small $\Delta \psi_i = \psi_i-\psi_{i+1}$, can be performed if the ring contains a large number of particles so that $\Delta \psi_i$ is small.
With this effective one-dimensional approximation, the dynamics reads: 
\begin{equation}
\label{eq:tangential_dynamics_second}
\dot{v}^t_i =-  K \left[2 r_i - r_{i+1}-r_{i-1} \right] - \gamma v^t_i + (f^a_i)^t +\sqrt{2\gamma T}\eta^t_i \,.
\end{equation}
To proceed further, it is convenient to switch from the ABP to the Ornstein-Uhlenbeck particle (AOUP) model~\cite{martin2020statistical, dabelow2019irreversibility, berthier2019glassy, maggi2017memory, wittmann2018effective}, approximating the active force of each particle, $f^a_i$, (in particular, its tangential component) with a one-dimensional Ornstein Uhlenbeck process:
\begin{equation}
\tau\dot{f}^a_i=-f^a_i + v_0\sqrt{2\tau} w_i
\end{equation}
where $\tau=1/D_r$ is the persistence time of the active force.
This strategy is particularly suitable to get analytical results both at the single-particle~\cite{szamel2014self, woillez2020nonlocal} and at the collective level~\cite{farage2015effective, fodor2016far, marconi2016velocity, wittmann2017effective, caprini2020time}. 
Indeed, AOUP and ABP active forces are characterized by the same temporal autocorrelation function~\cite{farage2015effective}. 
This ingredient seems to be crucial and, as a consequence, the AOUP can reproduce the main phenomenology experienced by the ABP model such as the accumulation near boundaries~\cite{caprini2018active, das2018confined, caprini2019activechiral} and the motility induced phase separation~\cite{fodor2016far, maggi2020universality}.
In particular, it has been recently employed to analytically predict the spatial profile of the velocity correlations in dense homogeneous systems of ABP~\cite{caprini2020hidden,caprini2020time}. 
The success of this approach has been corroborated, in Ref~\cite{caprini2020active}, by the direct comparison between the single-particle velocity distribution of ABP and AOUP at high density, which reveals a good agreement between the two models for a broad range of parameters.
For these reasons, we adopt the AOUP approximation to proceed further.

\section{Spatial velocity correlations in the Fourier space}\label{appendix:Fourier}

In this appendix, we derive the profile of the spatial velocity correlation in the Fourier space, given by Eq.~\eqref{eq:G0}.
The dynamics~\eqref{eq:tangential_dynamics} with the force~\eqref{eq:force_tangential} has the same structure as the equation of motion of the one-dimensional system (one-dimensional active particles on a line with periodic boundary conditions) studied in Ref.~\cite{caprini2020time}, upon replacing the position on the line with the position on the ring.
In particular, it is convenient to introduce the displacement of the $n$-th particle, $u_n=r_n - n \bar{r}$, from its positions $n \bar{r}$ on the ring and, then to evaluate Eq.~\eqref{eq:tangential_dynamics_second} in Fourier space. 
The discrete Fourier transforms of $u_j$, of the tangential velocity, $v_j$, and of the active force along the tangential direction, $f^a$, are defined as:
\begin{flalign}
\hat{u}(q)& = \frac{1}{N}\sum_{n=1}^N e^{-i n \cdot q} u_n \\
\hat{v}(q)& = \frac{1}{N}\sum_{n=1}^N e^{-i n \cdot q} v_n \\
\hat{f}^a(q)& = \frac{1}{N}\sum_{n=1}^N e^{-i n \cdot q} f^a_n \,,
\end{flalign}
where we have omitted the superscript $t$, for simplicity, and $q=  2\pi j/N$, with $j=0, 1, 2, ..., N-1$.
The dynamics in Fourier space assumes a simple form:
\begin{flalign}
\label{eq:tangential_dynamics_second}
\frac{d}{dt}\hat{u}(q) &= \hat{v}(q)\\
\frac{d}{dt}{\hat{v}(q)} &= - \omega^2(q) \hat{u}(q) - \gamma \hat{v}(q) + \hat{f}^a(q) +\sqrt{2\gamma T}\hat{\eta} \\
\frac{d}{dt}{\hat{f}^a(q)} &= -\frac{{\hat{f}^a(q)}}{\tau}  +  \sqrt{\frac{2}{\tau}}v_0\,\hat{\eta} \,.
\end{flalign}
where the frequency $\omega^2(q)$ reads:
$$
\omega^2(q) = 2 K\left[1-\cos(q)\right] \,.
$$
The crucial difference between the analysis on the ring and that on an infinite line (or with periodic boundary conditions) regards the infinite volume limit.
A ring of radius $\bar{R}$ can .accomodate  a maximal number of particles $N_m$  fixed by the density of the system.
This implies the presence of a physical lower cutoff in Fourier space. 
The one-dimensional theory developed in Ref.~\cite{caprini2020time}, can be easily adapted to the active underdamped case, following Ref.~\cite{caprini2020spatial}
The tangential velocity correlation function between different particles in Fourier space reads:
\begin{equation}
\label{eq:FourierTransform}
\langle \hat{v}(q) \hat{v}(-q) \rangle = \frac{T}{m} + \frac{f_0^2}{m} \frac{\tau}{\gamma}\frac{1}{1+\tau/\tau_i}\frac{1}{1+\frac{\tau^2}{1+\tau/\tau_I} \omega^2(q)}
\end{equation}
where we have omitted the superscript, $t$, for conciseness.
This profile Eq.~\eqref{eq:FourierTransform} coincides with Eq.~\eqref{eq:G0} if we neglect the first term taking the limit $T \ll f_0^2 \gamma^2=v_0^2$.
Equation~\eqref{eq:FourierTransform} is of the form:
$$
\langle \hat{v}(q) \hat{v}(-q) \rangle = A + \frac{B}{1+2\dfrac{\ell^2}{\bar{r}^2} (1-\cos{q})} \,,
$$
where
\begin{flalign}
&A=\frac{T}{m} \ll B, \\
&B=\frac{f_0^2}{m} \frac{\tau}{\gamma}\frac{1}{1+\tau/\tau_i}
\end{flalign}
and
$$
\ell^2=\bar{r}^2\frac{\tau^2}{1+\tau/\tau_I} \frac{K}{m} \,.
$$
The Fourier coefficient of this function are
\begin{equation}
\langle  v_{i+n}  v_i \rangle = A\delta_{n, 0} + B \sum_q  \frac{\cos{(qn)}}{1+2\dfrac{\ell^2}{\bar{r}^2}(1-\cos{q})}
\end{equation}
where $q=2\pi j/N $ with $j=0, ..., N-1$. 
We switch to the integral representation approximating the sum with an integral and subtracting the mode with $j=0$, we have:
\begin{equation}
\label{eq:fourier_integrals}
\langle \delta v_{i+n} \delta v_i \rangle = \frac{B}{\pi} \int_{2\pi/N}^{\pi}dq\,  \frac{\cos{(qn)}}{1+2\dfrac{\ell^2}{\bar{r}^2}(1-\cos{q})} \,,
\end{equation}
where $\delta v^t(r)= v^t(r) - \bar{v}^t$ and $\delta v^t(0)= v^t(0) - \bar{v}^t$, with $\bar{v}^t=1/N\sum_i v^t_i$, i.e. the spatial average. 
The integration limits of the integral provide the physical cutoff associated with the system.

\section{Small persistence regime, $\ell \ll L$}\label{appendix:smallpersistence}

In this appendix, we derive the spatial profile of $C_c(r)$ in the limit $\ell \ll L$, i.e. Eq.~\eqref{eq:exp_shap_C}.
Let us start from the infinite volume limit, which allows the approximation, $2\pi/N\to0$.
The discrete nature of $n$ allows us to solve the integrals~\eqref{eq:fourier_integrals} for every $n$.
Explicitly, the integral can be evaluated in terms of algebraic functions that can be calculated by introducing the variable:
$$
u^2=1+\frac{\bar{r}^2}{2\ell^2} = 1+\frac{m}{2K\frac{\tau^2}{1+\tau/\tau_I}} \,.
$$
Specifically, we get:
\begin{flalign}
g_1=\langle v_{i+1} v_i \rangle &=\frac{B}{2\ell^2} \left[ \frac{u^2}{\sqrt{u^4-1}} - 1  \right]=\\
&= \frac{B}{\ell^2} \left[ \frac{u^2-\sqrt{u^4-1}}{\sqrt{u^4-1}}   \right] \nonumber\\
g_2=\langle v_{i+2} v_i \rangle &=\frac{B}{ 2\ell^2}\left[ \frac{-1 +2u^4}{\sqrt{u^4-1}}  -2u^2\right]=\\
&=\frac{B}{ 2\ell^2}\frac{ \left(u^2- \sqrt{u^4-1}\right)^2 }{\sqrt{u^4-1}}\nonumber\\
... \nonumber\\
g_n=\langle v_{i+n} v_i \rangle &=\frac{B}{ 2\ell^2}\frac{ \left(u^2- \sqrt{u^4-1}\right)^n }{\sqrt{u^4-1}} \,.
\end{flalign}
It is convenient to express $g_n$ in terms of the variable $p=\bar{r}^2/(2\ell^2)$, obtaining: 
$$
g_n=\frac{B}{ 2\ell^2}\frac{ \left(1+p - \sqrt{(1+p)^2-1}\right)^n }{\sqrt{(1+p)^2-1}} \,.
$$
By performing a Taylor expansion for small $p$, holding in the regime of parameters $\frac{\tau ^2}{1+\tau/\tau_I} \gg m/K$, we have:
$$
g_n \propto\frac{\left( 1-\sqrt{2}\sqrt{p} \right)^n}{\sqrt{2}\sqrt{p}} \,.
$$
Now, taking formally the limit $n\to\infty$ (more physically $n\gg n(\sqrt{2 p})$ and thus $ 1/(\sqrt{2 p}) \gg 1$), we have:
$$
g_n \propto\frac{\left( 1-\frac{n\sqrt{2}\sqrt{p} }{n} \right)^n}{\sqrt{2}\sqrt{p}} \to \frac{e^{-n\sqrt{2p} }}{\sqrt{2}\sqrt{p}}\,,
$$
that leads to the exponential profile of the prediction~\eqref{eq:exp_shap_C} upon plugging the definition of $p$ in the above expression:
$$
C(r) \propto \exp{\left( - \frac{ n \bar{r}}{\ell}  \right)} \,,
$$
where
$$
\ell^2= \bar{r}^2 \frac{U''(\bar{r})}{m} \frac{\tau^2}{1+\frac{\tau}{\tau_I}} \,.
$$
We stress again that these results hold if we can consider the limit $N\to 0$ to perform the integral, a condition holding only if $\ell \ll L$.

\section{Large persistence regime, $\ell \gg L$}\label{appendix:largetau}

In this appendix, we derive the analytical predictions for $C_c(r)$ and $\lambda$ when $\ell \gg L$, namely Eqs.~\eqref{eq:Cc(r)_prediction} and~\eqref{eq:scalefree_lambda}.
To obtain these predictions one should be able to analytically calculate the integral~\eqref{eq:fourier_integrals} for every $n$ without assuming the condition $N\to\infty$. 
To get analytical results, we approximate the integral~\eqref{eq:fourier_integrals} as follows:
\begin{equation}
\begin{aligned}
\label{eq:approx_integral}
\langle \delta v_i \delta v_{i+n} \rangle \approx  & \frac{B}{2} \frac{\bar{r}^2}{\ell^2} \int_{2\pi/N}^{\pi}\frac{dq}{\pi}\,  \frac{\cos{(qn)}}{(1-\cos{q})}=\frac{B}{2} \frac{\bar{r}^2}{\ell^2} \mathcal{I}(n, N) 
\end{aligned}
\end{equation}
where we have assumed that $T \ll v_0^2$ (in such a way that $A \ll B$) and we can neglect the first term also for $n=0$.
To get the approximation~\eqref{eq:approx_integral}, we have also assumed that 
\begin{equation}
\frac{\ell^2}{\bar{r}^2}  \gg \frac{L^2}{4\pi^2}
\end{equation}
a condition that is fundamental to neglect the factor $1$ in the denominator of the integral~\eqref{eq:fourier_integrals}.
This approximation can be performed because of the lower cutoff on the integral, $2\pi/N$ that sets the minimal $q$ value.
Indeed, the integral~\eqref{eq:approx_integral} is divergent for $q\to 0$ at variance with the integral~\eqref{eq:fourier_integrals} that is always finite.
  
The integral~\eqref{eq:approx_integral} can be solved for generic $n$ in terms of a series of trigonometric functions and reads:
\begin{equation}
\mathcal{I}(0,N) = \frac{\cos(\pi/N)}{\pi \sin(\pi/N)}
\end{equation}
for $n=0$, while for generic $n>0$, we have:
\begin{equation}
\begin{aligned}
\mathcal{I}(n,N)&=\frac{n }{N} (2-N) + \frac{n}{\pi}\frac{\cos(\pi/N)}{\sin(\pi/N)} \\
&- \frac{n}{\pi\sin(\pi/N)}\sum_{j=1}^{n-1} \frac{1}{j(j+1)} \cos\left[(2j+1)\frac{\pi}{N}\right] \,.
\end{aligned}
\end{equation}
Assuming to deal with a large number of particles $N\sim 10^2, 10^3$ (as in the numerical work) the expression can be further simplified assuming that $N\gg1$:
\begin{equation}
\begin{aligned}
\mathcal{I}(n,N)&=\frac{n }{N} (2-N) + \frac{n}{\pi^2}N \\
&- \frac{n}{\pi^2}N\sum_{j=1}^{n-1} \frac{1}{j(j+1)} \cos\left[(2j+1)\frac{\pi}{N}\right] \,,
\end{aligned}
\end{equation}
where we have neglected orders $1/N$.
We can also rewrite the cosine as an infinite series: 
$$
\cos\left[(2j+1)\frac{\pi}{N}\right] = 1 + M[j] \,, 
$$
where 
$$
M[j]=\sum_{k=1}^{\infty} \frac{(-1)^k}{(2k)!} (2j+1)^{2k} \left(\frac{\pi}{N}\right)^{2k} \,.
$$
In this way, the normalized spatial velocity correlation reads:
\begin{equation}
\frac{\langle \delta v_i  \delta v_{i+n} \rangle}{\langle \delta v_i^2 \rangle} = \frac{\pi^2}{N}\mathcal{I}(n,N)
\end{equation}
since $ \mathcal{I}(0,N) \approx N/\pi^2$ (neglecting orders $1/N$) and, thus:
\begin{equation}
\begin{aligned}
\frac{\langle \delta v_i  \delta v_{i+n} \rangle}{\langle \delta v_i^2 \rangle} &= \frac{\pi^2 n}{N^2} (2-N) + n  - n \sum_{j=1}^{n-1} \frac{1 + M[j]}{j(j+1)} \\
&=1+\frac{\pi^2 n}{N^2}(2-N) - n\sum_{j=1}^{n-1} \frac{M[j]}{j(j+1)}
\end{aligned}
\end{equation}
where we have used that $\sum_{j=1}^{n-1} \frac{1}{j(j+1)} = 1-\frac{1}{n}$.
To proceed further, we note that the leading contributions in the remaining sum are those where $j$ appears at the maximal power in each of the infinite terms of the sum defining $M[j]$.
To fix the ideas, we evaluate the first two terms of the sum:
\begin{equation}
\begin{aligned}
-& n\sum_{j=1}^{n-1} \frac{M[j]}{j(j+1)} = -n \biggl( -\frac{1}{2}\frac{\pi^2}{N^2} \left[4n -3 -\frac{1}{n} \right] \\
&+ \frac{\pi^4}{4! N^4} \left[\frac{16}{3}n^3 + \frac{8}{3}n-\frac{21}{3}-\frac{1}{n} \right]\biggr) + ...\\
&\approx 2 \left(\frac{n}{N}\right)^2 \pi^2 - \frac{2}{9} \left(\frac{n}{N}\right)^4 \pi^4 \,,
\end{aligned}
\end{equation}
where we have neglected orders $n/N^2$ and higher orders (such as $1/N^2$, $(n/N^2)^2$, $n/N^4$ and $1/N^4$).
The other terms involved in the sum contain higher-order powers of the form $(n/N)^{\alpha}$, with $\alpha=6, 8, 10, ... $.
Plugging the results together, we have:
\begin{equation}
\label{app:eq_Ccn}
\frac{\langle \delta v_i  \delta v_{i+n} \rangle}{\langle \delta v_i^2 \rangle} \approx 1- \frac{\pi^2 n}{N} + 2 \left(\frac{n}{N}\right)^2 \pi^2 - \frac{2}{9} \left(\frac{n}{N}\right)^4 \pi^4 
\end{equation}
where we have just neglected orders $(n/N)^6$ and subleading orders $(n/N^2)$.
All the terms of  the orders $(n/N)^{k}$ can be summed together.
In particular, we get
\begin{equation}
\begin{aligned}
- n\sum_{j=1}^{n-1} &\frac{M[j]}{j(j+1)} \approx - \sum_{k=1}^{\infty} \frac{(-1)^k}{(2k)!} \left(\frac{2\pi n}{N}\right)^{2k}\frac{1}{2k-1}\\
&=-\left[ 1 - \cos\left(\frac{2\pi n}{N} \right) - \frac{2\pi n}{N} \text{SinInt}\left(\frac{2\pi n}{N}\right) \right]
\end{aligned}
\end{equation}
that is exact unless of the subleading order $n^{2k-1}/N^{2k}$.
We can easily observe that, by expanding the cosine and the Sinintegral function in powers of $n/N$, we get the correcting terms appearing in the profile of the spatial velocity correlation functions, i.e. Eq.~\eqref{app:eq_Ccn}.

Switching to a continuous notation such that $v_i \to v(r)$, being $r$ the coordinate along the ring, one obtains:
\begin{equation}
\label{eq:prediction_xllL}
\frac{\langle \delta v(r) \delta v(0) \rangle}{\langle \delta v^2 \rangle} = 1- \frac{r}{ \bar{r}}\frac{\pi^2}{N} + 2 \left(\frac{x}{\bar{r} N}\right)^2 \pi^2
\end{equation}
where $\bar{r}$ is the average distance between neighboring particles along the ring (and $v \equiv v_t$ is the tangential component of the particle velocity).
Eq.~\eqref{eq:prediction_xllL} corresponds to the prediction~\eqref{eq:Cc(r)_prediction} and the main correction occurs at the order $(r/L)^4$.

\section{Numerical study of the parameter $r_0$}

\begin{figure}[!t]
\centering
\includegraphics[width=0.95\linewidth,keepaspectratio]{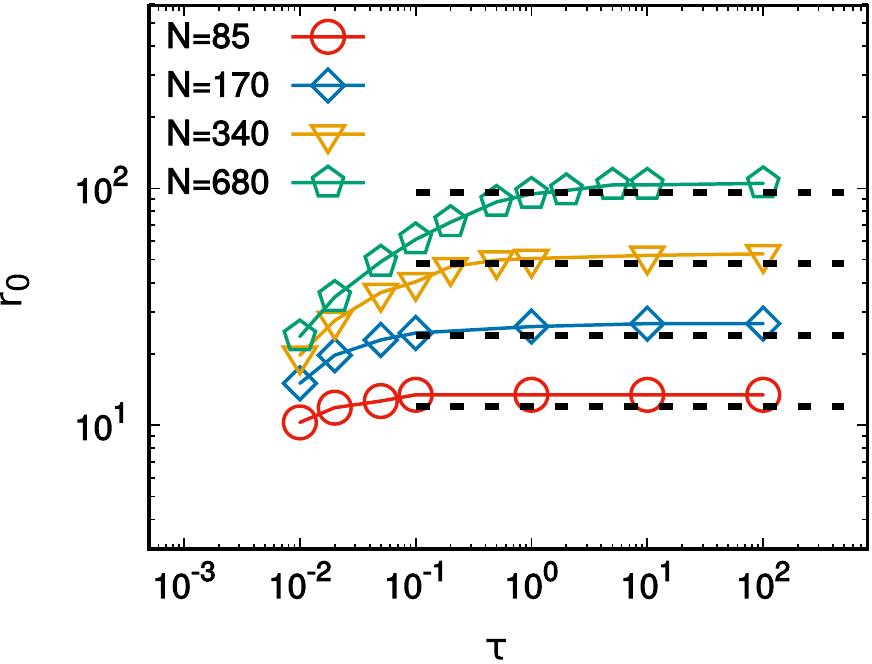}
\caption{\label{fig:r0_graph}
$r_0$ as a function of $\tau$ for different values of the system size $L=\bar{r}N$, as indicated in the legend.
The dashed black lines are marked in correspondence with the values predicted by Eq.~\eqref{eq:r0_vs_tau}.
The other parameters are $T=10^{-1}$, $\gamma=10^2$, $v_0=50$ and $\epsilon=10^2$.
}
\end{figure}

In this appendix, we study the parameter $r_0$ to check the relation~\eqref{eq:r0_vs_tau}, for completeness.
We remind that $r_0$ is defined as the distance at which the connected spatial correlation of the velocity (here, its tangential component with respect to the center of the ring) vanishes, $C_c(r_0)=0$.
Fig.~\eqref{fig:r0_graph} plots $r_0$ as a function of $\tau$ for different values of the system size, $L=\bar{r} N$.
This observable cannot be evaluated for small values of $\tau$ (for which the system is in the small persistence regime). Indeed, in that case, the correlation function has an exponential decay and does not reach negative values. Thus, the plot shows values such that $\tau\geq10^{-1}$.
Each curve (at fixed $L=\bar{r}N$) increases with $\tau$ until it saturates when the system enters the large persistence regime characterized by collective rotations. 
The value of the plateau, which is determined by the system size, is calculated using Eq.~\eqref{eq:r0_vs_tau} (see the comparison between colored points and dashed black lines in Fig.~\eqref{fig:r0_graph}) showing a fair agreement between data and the theoretical predictions for each system size.

\bibliographystyle{rsc} 

\bibliography{under.bib}

\end{document}